\documentclass[a4,12pt,epsf]{article} 

\usepackage{graphicx,amsmath} 
\usepackage{amsmath,amsfonts,latexsym,amssymb}
\usepackage{verbatim,amsthm,curves,graphics}
\usepackage{mathrsfs}

\def\ben{\begin{equation}}
\def\een{\end{equation}}
\def\bena{\begin{eqnarray}}
\def\eena{\end{eqnarray}}
\def\half{{1\over 2}}
\def\quater{{1 \over 4}}

\def\f(#1/#2){\frac{#1}{#2}} 
\def\Frac(#1/#2){\left(\frac{#1}{#2}\right)} 
\def\chris(#1-#2-#3){{\mit \Gamma}^{#1}{}_{{#2}{#3}} }
\def\tilchris(#1-#2-#3){\tilde{{\mit \Gamma}}^{#1}{}_{{#2}{#3}}}
\def\hatchris(#1-#2-#3){\hat{{\mit \Gamma}}^{#1}{}_{{#2}{#3}}}

\newcommand{\non}{\nonumber}
\theoremstyle{definition}

\newcommand{\dd}{{\rm d}}

\newcommand{\mr}{{\mathbb R}}

\newcommand{\e}{{\rm e}}
\renewcommand{\pounds}{{\mathscr L}}

\begin{document}

\title{A Higher Dimensional Stationary Rotating Black Hole Must be Axisymmetric} 

\author{
Stefan Hollands$^{1}$\thanks{\tt hollands@theorie.physik.uni-goe.de}\:,
Akihiro Ishibashi$^{2}$\thanks{\tt akihiro@midway.uchicago.edu}\:,
and Robert M. Wald$^{2}$\thanks{\tt rmwa@midway.uchicago.edu} 
\\ \\
{\it ${}^{1}$Institut f\"ur Theoretische Physik, 
     Universit\"at G\" ottingen,} \\
{\it D-37077 G\" ottingen, Germany,} \medskip \\ 
{\it ${}^{2}$Enrico Fermi Institute and Department of Physics, } \\
{\it The University of Chicago, Chicago, IL 60637, USA} \\ 
    }

\maketitle

\abstract{A key result in the proof of black hole uniqueness in
$4$-dimensions is that a stationary black hole that is
``rotating''---i.e., is such that the stationary Killing field is not
everywhere normal to the horizon---must be axisymmetric. The proof of
this result in $4$-dimensions relies on the fact that the orbits of
the stationary Killing field on the horizon have the property that
they must return to the same null geodesic generator of the horizon
after a certain period, $P$. This latter property follows, in turn,
from the fact that the cross-sections of the horizon are
two-dimensional spheres. However, in spacetimes of dimension greater
than $4$, it is no longer true that the orbits of the stationary
Killing field on the horizon must return to the same null geodesic
generator. In this paper, we prove that, nevertheless, a higher
dimensional stationary black hole that is rotating must be
axisymmetric. No assumptions are made concerning the topology of the
horizon cross-sections other than that they are compact. However, 
we assume that the horizon is non-degenerate and, 
as in the $4$-dimensional proof, that the spacetime is analytic.} 

\section{Introduction} \label{sect:1}

Consider an $n$-dimensional stationary spacetime containing a black
hole. Since the event horizon of the black hole must be mapped into
itself by the action of any isometry, the asymptotically timelike
Killing field $t^a$ must be tangent to the horizon. Therefore, we have
two cases to consider: (i) $t^a$ is normal to the horizon, i.e., tangent
to the null geodesic generators of the horizon; (ii) $t^a$ is not normal
to the horizon. In $4$-dimensions it is known that in case (i), 
for suitably regular 
non-extremal vacuum or Einstein-Maxwell black holes, the black hole 
must be static~\cite{SW92,CW94}. 
Furthermore, in $4$-dimensions it is 
known that in case (ii), under fairly general assumptions about 
the nature of the matter content but assuming 
analyticity of the spacetime and non-extremality of the black hole,
there must exist an additional Killing field that is normal to the
horizon. It can then be shown that the black hole must be 
axisymmetric\footnote { 
In this paper, by ``axisymmetric'' we mean that spacetime 
possesses one-parameter group of isometries isomorphic to $U(1)$ whose 
orbits are spacelike. We do not require that the Killing field vanishes 
on an ``axis.''
} 
as well as stationary~\cite{H72,HE}. 
This latter result is often referred to as a ``rigidity theorem,'' 
since it implies that the horizon generators of a ``rotating'' 
black hole (i.e., a black hole for which $t^a$ is not normal 
to the horizon) must rotate rigidly with respect to infinity. 
A proof of the rigidity theorem in $4$-dimensions which partially 
eliminates the analyticity assumption was given by Friedrich, Racz,
and Wald~\cite{FRW99,Racz00}, based upon an argument of Isenberg and
Moncrief~\cite{MI83,IM85} concerning the properties of spacetimes with
a compact null surface with closed generators. The above results for
both cases (i) and (ii) are critical steps in the proofs of black hole
uniqueness in $4$-dimensions, since they allow one to apply Israel's
theorems~\cite{Israel67,Israel68} 
in case (i) and the Carter-Robinson-Mazur-Bunting 
theorems~\cite{Carter71,Robinson75,Mazur82,Bunting83} in case (ii). 

Many attempts to unify the forces and equations of nature involve the
consideration of spacetimes with $n>4$ dimensions. Therefore, it is of
considerable interest to consider a generalization of the rigidity
theorem to higher dimensions, especially in view of the fact that
there seems to be a larger variety of black hole solutions (see e.g.,
~\cite{ER02b,GHP02,GLPP04}), the classification of which has not been
achieved yet. 
\footnote{ 
There have recently appeared several works on general properties 
of a class of stationary, axisymmetric vacuum solutions, 
including an $n$-dimensional generalization of the Weyl solutions 
for the static case (see 
e.g.,~\cite{CG04,ER02a,Harmark04,Harmark-Olesen05}, and see 
also~\cite{Mishima-Iguchi05,Tomizawa-etal05} and references therein 
for some techniques of generating such solutions in $5$-dimensions).  
} 
The purpose of this paper is to present a proof of the
rigidity theorem in higher dimensions for non-extremal black holes.

The dimensionality of the spacetime enters the proof of the rigidity
theorem in $4$-dimensions in the following key way: 
The expansion and
shear of the null geodesic generators of the horizon of a stationary
black hole can be shown to vanish (see below).
The induced (degenerate) metric on the
$(n-1)$-dimensional horizon gives rise to a Riemannian metric,
$\gamma_{ab}$, on an arbitrary 
$(n-2)$-dimensional cross-section, $\Sigma$, of the horizon.
On account of the vanishing
shear and expansion, all cross-sections of the horizon are isometric,
and the projection of 
the stationary Killing field $t^a$ onto $\Sigma$ gives rise to
a Killing field, $s^a$, of $\gamma_{ab}$ on $\Sigma$. In case (ii), $s^a$
does not vanish identically.  Now, when $n=4$, it is known that
$\Sigma$ must have the topology of a $2$-sphere, $S^2$. Since the
Euler characteristic of $S^2$ is nonzero, it follows that $s^a$
must vanish at some point $p \in \Sigma$. However, since $\Sigma$
is $2$-dimensional, it then follows that the isometries generated by
$s^a$ simply rotate the tangent space at $p$. It then follows that 
all of the orbits
of $s^a$ are periodic
with a fixed period $P$, from which it follows that, after
period $P$, the orbits of $t^a$ on the horizon must return to the
same generator. Consequently, if we identify points in spacetime
that differ by the action of the stationary isometry of parameter $P$,
the horizon becomes a compact null surface with closed null geodesic
generators. The theorem of Isenberg and Moncrief~\cite{MI83,IM85} then
provides the desired additional Killing field normal to this null
surface.

In $n>4$ dimensions, the Euler characteristic of $\Sigma$ may vanish,
and, even if it is non-vanishing, if $n>5$ there is no reason that the
isometries generated by $s^a$ need have closed orbits even when $s^a$
vanishes at some point $p \in \Sigma$. Thus, for example, even in the
$5$-dimensional Myers-Perry black hole solution~\cite{MP86} with cross
section topology $\Sigma = S^3$, one can choose the rotational
parameters of the solution so that the orbits 
of the stationary Killing field $t^a$ do
not map horizon generators into themselves.

One possible approach to generalizing the rigidity theorem to higher
dimensions would be to choose an arbitrary $P > 0$ and identify points
in the spacetime that differ by the action of the stationary isometry
of parameter $P$. Under this identification, the horizon would again
become a compact null surface, but now its null geodesic generators
would no longer be closed. The rigidity theorem would follow if the results
of~\cite{MI83,IM85} could be generalized to the case of compact null
surfaces that are ruled by non-closed generators. We have learned that
Isenberg and Moncrief are presently working on 
such a generalization \cite{IM06},
so it is possible that the rigidity theorem can be proven in this
way. 

However, we shall not proceed in this manner, but rather will parallel
the steps of~\cite{MI83,IM85}, replacing arguments that rely on the
presence of closed null generators with arguments that rely on the
presence of stationary isometries. Since on the horizon we may write
\ben 
t^a = n^a + s^a \, , 
\een 
where $n^a$ is tangent to the null geodesic generators and $s^a$ is 
tangent to cross-sections of the horizon, the stationarity in essence
allows us to replace Lie derivatives with respect to $n^a$ by Lie derivatives
with respect to $s^a$. Thus, equations in~\cite{MI83,IM85} that can be solved 
by integrating quantities along the orbits of the closed null geodesics
correspond here to equations that can be solved if one can suitably integrate
these equations along the orbits of $s^a$ in $\Sigma$. Although the 
orbits of $s^a$ are not closed in general, 
we can appeal to basic results of ergodic theory 
together with the fact that
$s^a$ generates isometries of $\Sigma$ to solve these equations.

For simplicity, we will focus attention on the vacuum Einstein's equation,
but we will indicate in section~\ref{sect:4} 
how our proofs can be extended to models with a cosmological constant 
and a Maxwell field. 
As in~\cite{H72,HE} and in~\cite{MI83,IM85}, we will assume analyticity,
but we shall indicate how this assumption can be partially removed (to
prove existence of a Killing field {\it inside} the black hole) by
arguments similar to those given in~\cite{FRW99,Racz00}. The non-extremality
condition is used for certain constructions in the proof (as well
as in the arguments partially removing the analyticity condition), and it
would not appear to be straightforward to generalize our arguments
to remove this restriction when the orbits of $s^a$ are not closed.

\medskip

Our signature convention for $g_{ab}$ is $(-,+,+,\cdots)$. We define
the Riemann tensor by $R_{abc}{}^d k_d = 2\nabla_{[a} \nabla_{b]} k_c$ 
and the Ricci tensor by $R_{ab} = R_{acb}{}^c$. We also set $8\pi G = 1$. 

\section{Proof of existence of a horizon Killing field} 
\label{sect:2}

Let $(M,g_{ab})$ be an $n$-dimensional, smooth,
asymptotically flat, stationary solution
to the vacuum Einstein equation containing a black hole.
Thus, we assume the existence in the spacetime of a  
Killing field $t^a$ with complete orbits which are timelike near infinity.
Let $H$ denote a connected component of the portion of 
the event horizon of the black hole that lies to the 
future of ${\mathscr I}^-$. We assume that $H$ has 
topology $\mr \times \Sigma$, where $\Sigma$ is compact. 
Following Isenberg and Moncrief~\cite{MI83,IM85},
our aim in this section is to prove that there exists a 
vector field $K^a$ defined in a neighborhood of $H$ which is normal to $H$
and on $H$ satisfies 
\ben
\label{civ1}
\underbrace{\pounds_\ell \, \pounds_\ell \, \cdots \,
\pounds_\ell}_{m \,\, {\rm times}} 
(\pounds_K g_{ab}) = 0, \quad m=0,1,2, \dots, 
\een
where $\ell$ is an arbitrary vector field
transverse to $H$. As we shall show at the end of this section,
if we assume analyticity of $g_{ab}$ and of $H$ 
it follows that $K^a$ is a 
Killing field. We also will explain at the end of this section 
how to partially 
remove the assumption of analyticity of $g_{ab}$ and $H$. 

We shall proceed by constructing a candidate Killing field, $K^a$, and then
proving that eq.~(\ref{civ1}) holds
for $K^a$. This candidate Killing field is expected
to satisfy the following properties: (i) $K^a$ should be normal to $H$. 
(ii) If we define $S^a$ by
\ben
S^a = t^a - K^a
\een
then, on $H$, $S^a$ should be tangent to cross-sections\footnote{Note that
as already mentioned above, since $H$ is mapped into itself by the time
translation isometries, $t^a$ must be tangent to 
$H$, so $S^a$ is automatically tangent to $H$. Condition (iii) requires that
there exist a foliation of $H$ by cross-sections $\Sigma(u)$ such that 
each orbit of $S^a$ is contained in a single cross-section.} of $H$. 
(iii) $K^a$ should commute with $t^a$.
(iv) $K^a$ should have constant surface gravity on $H$, i.e., on $H$ we should
have $K^a \nabla_a K^b = \kappa K^b$ with $\kappa$ constant on $H$, 
since, by the zeroth law of black hole
mechanics, this property is known to hold
on any Killing horizon in any vacuum solution of Einstein's equation.

We begin by choosing a cross-section $\Sigma$, of $H$.
By arguments similar to those given in the proof of proposition 4.1 
of \cite{CW94}, we may assume without loss of generality that $\Sigma$
has been chosen so that each orbit of $t^a$ on $H$ intersects $\Sigma$
at precisely one point, so that $t^a$ is everywhere transverse to 
$\Sigma$. We extend $\Sigma$ 
to a foliation, $\Sigma(u)$, of $H$ by the action of 
the time translation isometries, i.e.,
we define $\Sigma(u) = \phi_u (\Sigma)$, where $\phi_u$ denotes the 
one-parameter group of isometries generated by $t^a$. 
Note that the function $u$ on $H$ that labels the cross-sections in this
foliation automatically satisfies 
\ben
\pounds_t u = 1 \, .
\een
Next, we define $n^a$ and $s^a$ on $H$ by 
\ben
t^a = n^a + s^a \, , 
\een
where $n^a$ is normal to $H$ and $s^a$ is tangent to $\Sigma(u)$. 
It follows from the transversality of $t^a$ that $n^a$ is everywhere
nonvanishing and future-directed.
Note also that $\pounds_n u = 1$ on $H$.
Our strategy is to extend this 
definition of $n^a$ to a neighborhood of $H$
via Gaussian null coordinates.
This construction of $n^a$ obviously
satisfies conditions (i) and (ii) above, and it also will be shown below
that it satisfies condition (iii). However,
it will, in general, fail to satisfy (iv). We shall then modify
our foliation so as to produce a new foliation $\tilde{\Sigma} (\tilde{u})$ 
so that (iv)
holds as well. We will then show that the corresponding 
$K^a=\tilde{n}^a$ satisfies 
eq.~(\ref{civ1}).

Given our choice of $\Sigma(u)$ and the corresponding choice of $n^a$ on $H$,
we can uniquely define 
a past-directed null vector field $\ell^a$ on $H$ by the requirements 
that $n^a \ell_a = 1$, and that $\ell^a$ is orthogonal 
to each $\Sigma(u)$. 
Let $r$ denote the affine parameter on the null geodesics determined by
$\ell^a$, with $r=0$ on $H$.
Let $x^A=(x^1, x^2, \dots, x^{n-2})$ 
be local coordinates on an open subset of $\Sigma$. 
Of course, it will take more than one coordinate patch to cover
$\Sigma$, but 
there is no problem in patching together local results, 
so no harm is done in pretending that $x^A$ covers $\Sigma$. 
We extend the coordinates $x^A$ from $\Sigma$ to $H$ 
by demanding that they be constant 
along the orbits of $n^a$. We then extend $u$ and $x^A$ to a neighborhood
of $H$ by requiring these quantities to be constant along the orbits
of $\ell^a$. It is easily seen that the quantities $(u,r, x^A)$ define
coordinates covering a neighborhood of $H$. Coordinates that are constructed
in this manner are known as {\it Gaussian null coordinates} and are unique
up to the choice of $\Sigma$ and the choice of coordinates $x^A$ on $\Sigma$.
It follows immediately that on $H$ we have 
\ben
n^a = \left(\frac{\partial}{\partial u} \right)^a \,, 
\quad \ell^a = \left(\frac{\partial}{\partial r}\right)^a \,,
\een
and we extend $n^a$ and $\ell^a$ to a neighborhood of $H$ by these formulas.
Clearly, $n^a$ and $\ell^a$ commute, since they are coordinate vector fields.

Note that we have
\ben
\ell^a \nabla_a (n_b \ell^b) = \ell^b \ell^a \nabla_a n_b 
= \ell^b n^a \nabla_a \ell_b = \frac{1}{2} n^a \nabla_a (\ell^b
\ell_b) = 0 \, , 
\een
so $n_a \ell^a = 1$ everywhere, not just on $H$. Similarly, we have
$\ell_a (\partial/\partial x^A)^a = 0$ everywhere. It follows that in
Gaussian null coordinates, the metric in a neighborhood of $H$
takes the form
\ben
\label{gaussian}
g_{\mu \nu}\dd x^\mu \dd x^\nu 
= 2 \left( \dd r - r \alpha \dd u - r \beta_{A} \dd x^A \right) \dd u 
         + \gamma_{AB} \dd x^A \dd x^B \,,  
\een
where, again, $A$ is a labeling index that runs from  $1$ to $n-2$. 
We write
\ben
\beta_a = \beta_A (\dd x^A)_a \,, \,\,\,\, 
\gamma_{ab} = \gamma_{AB} (\dd x^A)_a (\dd x^B)_b \,.
\een
Note that $\alpha$, $\beta_a$, and $\gamma_{ab}$ are independent of 
the choice of coordinates, $x^A$, and thus are globally defined in 
an open neighborhood of $H$. 
{}From the form of the metric,
we clearly have $\beta_a n^a = \beta_a \ell^a = 0$ and 
$\gamma_{ab} n^a = \gamma_{ab} \ell^a = 0$. It then follows that $\gamma^a{}_b$
is the orthogonal
projector onto the subspace of the tangent space perpendicular 
to $n^a$ and $\ell^a$, where here and elsewhere, all indices 
are raised and lowered with the spacetime metric $g_{ab}$. Note that 
when $r \neq 0$, i.e., off of the horizon, $\gamma_{ab}$
differs from the metric $q_{ab}$, on the $(n-2)$-dimensional
submanifolds, $\Sigma(u,r)$, of constant $(u,r)$, since 
$n^a$ fails to be perpendicular to these surfaces. Here, $q_{ab}$ is defined
by the condition that $q^a{}_b$ is the orthogonal projector onto 
the subspace of the tangent space that is tangent to $\Sigma(u,r)$;
the relationship between $\gamma_{ab}$ and $q_{ab}$ is given by
\ben
q_{ab} = r^2 \beta^c \beta_c \ell_a \ell_b - 2r \beta_{(a} \ell_{b)} 
+ \gamma_{ab} \, .
\een
However, since on $H$ (where $r=0$), we have 
$\gamma_{ab} = q_{ab}$, we will refer to $\gamma_{ab}$
as the metric on the cross-sections $\Sigma(u)$ of $H$.

Thus, we see that in Gaussian null coordinates
the spacetime metric, $g_{ab}$, is characterized by
the quantities $\alpha$, $\beta_a$, and $\gamma_{ab}$. 
In terms of these quantities, 
if we choose $K^a = n^a$, then the 
condition~\eqref{civ1} will hold if and only if the conditions
\bena
\label{cid}
\underbrace{\pounds_\ell \, \pounds_\ell \, \cdots \,
\pounds_\ell}_{m \,\, {\rm times}} 
\left(\pounds_n \gamma_{ab} \right) &=& 0 \, ,
\non\\
\underbrace{\pounds_\ell \, \pounds_\ell \, \cdots \,
\pounds_\ell}_{m \,\, {\rm times}} 
\left(\pounds_n \alpha \right) &=& 0 \, ,
\non\\ 
\underbrace{\pounds_\ell \, \pounds_\ell \, \cdots \,
\pounds_\ell}_{m \,\, {\rm times}} 
\left(\pounds_n \beta_a \right) &=& 0 \, ,
\eena
hold on $H$.

Since the vector fields $n^a$ and $\ell^a$ are uniquely determined by
the foliation
$\Sigma(u)$ and since $\phi_{u}[\Sigma(u')] = \Sigma(u+u')$
(i.e., the time translations leave the foliation invariant), it follows 
immediately that $n^a$ and $\ell^a$ are invariant under $\phi_u$. 
Hence, we have $\pounds_t n^a = \pounds_t \ell^a = 0$, so, in 
particular, condition (iii) holds, as claimed above. 
Similarly, we have
$\pounds_t r = 0$ and $\pounds_t u = 1$ throughout the region where the
Gaussian null coordinates are defined. Since $\pounds_t g_{ab} = 0$, we
obtain from eq.~(\ref{gaussian})
\ben
0 = -2r \pounds_t \alpha \, \nabla_a u \, \nabla_b u - 2r \, \pounds_t
\beta_{(a} 
\nabla_{b)} u
+ \pounds_t \gamma_{ab} \, .
\een
Contraction of this equation with $n^a n^b$ yields
\ben
\pounds_t \alpha = 0 \, .
\label{ta}
\een
Contraction with $n^a$ then yields
\ben
\pounds_t \beta_a = 0 \, ,
\label{tb}
\een
and we then also immediately obtain
\ben
\pounds_t \gamma_{ab} = 0 \, .
\label{tc}
\een

The next step in the analysis is to use the Einstein 
equation $R_{ab} n^a n^b = 0$ on $H$, in a manner completely in
parallel with the 
4-dimensional case~\cite{HE}. This equation 
is precisely the Raychaudhuri
equation for the congruence of null curves defined by 
$n^a$ on $H$. Since that congruence is twist-free on $H$, we obtain on $H$
\ben
\frac{\dd}{\dd \lambda} \theta = -\frac{1}{n-2} \theta^2 - 
\widehat \sigma_{ab} \widehat \sigma^{ab}\, , 
\een 
where $\theta$ denotes the expansion of the null geodesic generators of
$H$, $\widehat \sigma_{ab}$ denotes their shear,
and $\lambda$ is the affine parameter along null geodesic generators of $H$
with tangent $n^a$. 
Now, by the same arguments as used to prove the area theorem~\cite{HE},
we cannot have $\theta < 0$ on $H$. 
On the other hand, 
the rate of change of the area, $A(u)$, of $\Sigma(u)$ (defined 
with respect to the metric $q_{ab} = \gamma_{ab}$) is given by
\ben 
\frac{\dd }{\dd u} A(u) = 
\half \int_\Sigma \left( 
                         \frac{\partial \lambda}{\partial u} 
                  \right) \theta \sqrt{\gamma} \, \dd^{n-2} x \, . 
\een
However, since $\Sigma(u)$ is related to $\Sigma$ by the isometry
$\phi_u$, the left side of this equation must vanish.
Since 
$\partial \lambda/ \partial u > 0$ on $H$,  
this shows that $\theta = 0$ on $H$. It then follows immediately
that $\widehat \sigma_{ab}=0$ on $H$. Now on $H$, 
the shear is equal to 
the trace free part of $\pounds_n\gamma_{ab}$ 
while the expansion is equal to the trace of this quantity. 
So we have shown that $\pounds_n \gamma_{ab} = 0$ on $H$. Thus,
the first equation in eq.~(\ref{cid}) holds with $m=0$.

However, $n^a$ in general fails to satisfy condition (iv) above. Indeed,
from the
form, eq.~(\ref{gaussian}), of the metric, we see that the surface gravity,
$\kappa$, associated with $n^a$ is simply $\alpha$, and there is no reason
why $\alpha$ need be constant on $H$. Since $\pounds_n \gamma_{ab} = 0$ on $H$,
the Einstein equation $R_{ab} n^a (\partial/\partial x^A)^b = 0$ on $H$ yields
\ben
 D_a \alpha = \frac{1}{2}  \pounds_n \beta_a \,, 
\label{dalpha}
\een
(see eq.~(\ref{nAR}) of Appendix~\ref{sect:A}) where $D_a$ denotes 
the derivative operator on $\Sigma(u)$, i.e., 
$D_a \alpha = {q_a}^b \nabla_b \alpha = {\gamma_a}^b \nabla_b \alpha$.
Thus, if $\alpha$ is not constant on $H$, then the last
equation in eq.~(\ref{cid}) fails to hold even when $m=0$.
As previously indicated, our strategy is repair this problem by choosing
a new cross-section $\tilde{\Sigma}$ so that the corresponding 
$\tilde{n}^a$ arising from the Gaussian null coordinate 
construction will have constant surface gravity on $H$. 
The determination of this $\tilde{\Sigma}$ requires some intermediate 
constructions, to which we now turn. 

First, since we already know that $\pounds_t \gamma_{ab} = 0$
everywhere and that
$\pounds_n \gamma_{ab} = 0$ on $H$, it follows immediately from the fact
that $t^a=s^a+n^a$ on $H$ that 
\ben
\pounds_s \gamma_{ab} = 0 \,,
\een
on $H$ (for any choice $\Sigma$). Thus, $s^a$ is a Killing 
vector field for the Riemannian metric 
$\gamma_{ab} = q_{ab}$ on $\Sigma$. 
Therefore the flow, $\hat \phi_\tau: \Sigma \to \Sigma$ 
of $s^a$ yields
a one-parameter group of isometries of $\gamma_{ab}$, which coincides 
with the projection of the flow $\phi_u$ of the original Killing 
field $t^a$ to $\Sigma$.

We define $\kappa$ to be the mean value 
of $\alpha$ on $\Sigma$, 
\ben\label{kdef}
 \kappa = \frac{1}{A(\Sigma)} \int_\Sigma \alpha \sqrt{\gamma} \, \dd^{n-2} x 
\,, 
\een
where $A(\Sigma)$ is the area of $\Sigma$ with respect to the 
metric $\gamma_{ab}$. 
In the following we will assume that $\kappa \neq 0$, i.e., that we are in 
the ``non-degenerate case." Given that $\kappa \neq 0$, we may assume 
without loss of generality, that $\kappa>0$.

We seek a new Gaussian null coordinate system $(\tilde{u}, 
\tilde{r}, \tilde{x}^A)$ satisfying all of the above properties of $(u,r,x^A)$
together with the additional requirement that $\tilde{\alpha} = \kappa$, i.e.,
constancy of the surface gravity. We now determine the conditions that 
these new coordinates would have to satisfy.
Since clearly $\tilde{n}^a$ must be 
proportional to $n^a$, we have
\ben
\label{nfn}
\tilde n^a = f n^a \,,  
\een
for some positive function $f$. Since $\pounds_t  \tilde{n}^a
= \pounds_t n^a = 0$, we must have $\pounds_t f = 0$. Since on $H$ we have
$n^a \nabla_a n^b = \alpha n^b$ and $\tilde{\alpha}$ is given by
\ben
\tilde n^a \nabla_a \tilde n^b = \tilde \alpha \tilde n^b \,,  
\een
we find that $f$ must satisfy
\ben
\tilde \alpha = \pounds_n f + \alpha f = -\pounds_s f + \alpha f = \kappa \,.
\label{f}
\een
The last equality provides an equation that must be satisfied by $f$ on
$\Sigma$.
In order to establish that a solution to this
equation exists, we first prove the
following lemma:

\medskip
\noindent
\paragraph{Lemma 1} 
For any $x \in \Sigma$, we have 
\ben
 \kappa = \lim_{S \to \infty} \frac{1}{S} 
 \int_0^S \alpha(\hat{\phi}_\tau(x)) \, \dd \tau \, .
\label{avgalpha}
\een
Furthermore, the convergence of the limit is uniform in $x$.
Similarly, $x$-derivatives of 
$S^{-1} \int_0^S \alpha(\hat{\phi}_\tau(x)) \, \dd \tau$ 
converge to $0$ uniformly in $x$ as $S \rightarrow \infty$.

\medskip
\noindent
{\bf Proof}: The von Neumann ergodic theorem (see e.g.,\cite{Walter})
states that if $F$ is an $L^p$ function 
for $1 \le p < \infty$ on a measure space $(X,\dd m)$ with finite measure, and
if $T_\tau$ is a continuous one-parameter group of measure preserving 
transformations on $X$, then 
\bena
\label{vnet}
F^*(x) &=& \lim_{S \to \infty} \frac{1}{S} \int_0^S F(T_\tau(x)) \, \dd \tau\\
&=& \lim_{\epsilon \to 0+} \epsilon \int_0^\infty \e^{-\epsilon \tau} 
F(T_\tau(x)) \, \dd \tau
\eena
converges in the sense of $L^p$ (and in particular almost everywhere).
We apply this theorem to $X = \Sigma$, $\dd m = \sqrt{\gamma} \, \dd^{n-2}x$, 
$F = \alpha$, and $T_\tau = \hat{\phi}_\tau$, to conclude that there is 
an $L^p$ function $\alpha^*(x)$ on $\Sigma$ 
to which the limit in the lemma converges. 
We would like to prove that $\alpha^*(x)$ is constant. To prove this,
we note that eq.~(\ref{dalpha}) together with the facts that
$\pounds_t \beta_a = 0$ and 
$t^a = n^a + s^a$ yields
\ben
D_b \alpha = -\half \pounds_s \beta_b \, .
\label{dalpha2}
\een
Now let
\ben
a(x,S) = \int_0^S \alpha(\hat{\phi}_\tau(x)) \, \dd \tau \, .
\een
Then 
\ben
D_b a(x,S) = -\half \left\{ \hat \phi_S^* \beta_b(x)  - \beta_b(x) \right\} \, , 
\een
and thus 
\bena
\label{estim0}
|a(x,S) - a(y,S)| &\le& C' \, \sup \{ [D^b a D_b a(z,S)]^{1/2};
\,\, z \in \Sigma \} \nonumber \\
&\le&  C' \, \sup \{ [\beta^b \beta_b(z)]^{1/2}; \,\,  z
 \in \Sigma\} = C < \infty \, , 
\eena
where $C,C'$ are constants independent of $S$ and $x$, and where
$C$ is finite because $\Sigma$ is compact. 
Consequently, $|a(x,S)-a(y,S)|$ is uniformly bounded in $S \ge 0$ and in
$x,y
\in \Sigma$. Thus, for all $x,y \in \Sigma$, we have
\ben
\label{estim}
\lim_{S \to \infty} \frac{1}{S} |
  a(x,S) - a(y,S)| \le 
 \lim_{S \to \infty} \frac{C}{S} = 0 \,. 
\een
Let $y \in \Sigma$ be such that
$a(y,S)/S$ converges as $S \rightarrow \infty$. (As already noted above,
existence of such a $y$ is guaranteed by the von Neumann ergodic 
theorem.) The above equation then
shows that, in fact, $a(x,S)/S$ must
converge for all $x\in \Sigma$ as $S \rightarrow \infty$ and that, 
furthermore, the limit is independent of $x$, as we desired to show.
Thus, $\alpha^*(x)$ is constant, and hence equal 
to its spatial average, $\kappa$. The estimate~\eqref{estim0} also
shows that the limit~(\ref{avgalpha}) is uniform in $x$. Similar
estimates can easily be obtained for the norm with
respect to $\gamma_{ab}$ of 
$[D_{c_1} \cdots D_{c_k} a(x,S) -  D_{c_1} \cdots D_{c_k} a(y,S)]$, for
any $k$. These estimates show that $x$-derivatives of $a(x,S)/S$ converge 
to $0$ uniformly in $x$. 
$\Box$ \hfill

\bigskip

We now are in a position to prove the existence of
a positive function $f$ on $\Sigma$
satisfying the last equality in eq.~(\ref{f}) on $\Sigma$. Let
\ben
\label{fdef}
f(x) = \kappa \int_0^\infty p(x,\sigma) \, \dd \sigma, 
\een
where $p(x, \sigma)>0$ is the function on $\Sigma \times \mr$ defined by
\ben\label{pdef}
p(x, \sigma) = {\rm exp} \left( - \int_0^\sigma \alpha (\hat{\phi}_{\tau}(x)) \, \dd \tau 
\right) \, .
\een
The function $f$ is well defined for almost all 
$x$ because $p(x, \sigma) < \e^{- \sigma(\kappa -\epsilon)}$ 
for any $\epsilon$ and sufficiently large $\sigma$, 
by Lemma~1. It also follows from the uniformity statement in 
Lemma~1 that $f$ is smooth on $\Sigma$.
By a direct calculation, using Lemma~1, we find that 
$f$ satisfies
\ben
\label{fdiff}
 - \pounds_s f(x) + \alpha(x) f(x) = \kappa \, ,
\een
as we desired to show.

We now can deduce how to choose the 
desired new Gaussian null coordinates. The new
coordinate $\tilde{u}$ must satisfy
\ben
\label{normali}
\pounds_t \tilde u = 1 \, , 
\een
as before. However, in view of eq.~(\ref{nfn}), it also must satisfy
\ben
\pounds_n \tilde u = n^a \nabla_a \tilde{u} 
= \frac{1}{f} \tilde{n}^a \nabla_a \tilde{u} = \frac{1}{f} \,.  
\een
Since $n^a = t^a - s^a$, we find that on $\Sigma$, $\tilde{u}$ must satisfy
\ben
1 - \pounds_s \tilde u = \frac{1}{f}  \,.  
\een
Substituting from eq.~(\ref{fdiff}), we obtain
\ben
\pounds_s \tilde u = 1 + \frac{1}{\kappa} (\pounds_s \ln f - \alpha)\,.  
\label{tildeu}
\een
Thus, if our new Gaussian null coordinates exist, there must exist a 
smooth solution to this equation. That this is the case is proven in the 
following lemma.

\medskip
\noindent
\paragraph{Lemma 2}
There exists a smooth 
solution $h$ to the following differential equation on $\Sigma$: 
\ben
\label{psh}
\pounds_s h(x) = \alpha(x) - \kappa \, .
\een

\medskip 
\noindent 
{\bf Proof}: First note that the orbit average of any function of the 
form $\pounds_s h(x)$ where $h$ is smooth
must vanish, so there could not possibly exist
a smooth solution to the above equation 
unless the average of $\alpha$ over any orbit is equal to
$\kappa$. However, this was
proven to hold in Lemma~1.
In order to get a solution to the above equation, 
choose $\epsilon>0$, 
and consider the regulated expression defined by
\ben
h_\epsilon(x) = - \int_0^\infty {\rm e}^{-\epsilon \tau} 
 \left[\alpha(\hat{\phi}_\tau(x)) - \kappa \right]\,\dd\tau \,. 
\label{hedef}
\een
Due to the exponential damping, this quantity is smooth, and satisfies
the differential equation
\ben
\pounds_s h_\epsilon(x) = \alpha(x) - \kappa - \epsilon h_\epsilon(x) \, .
\een
We would now like to take the limit as $\epsilon \to 0$ to get
a solution to the desired equation. However, it is not possible to 
straightforwardly take the limit as $\epsilon \to 0$ of $h_\epsilon(x)$, 
for there is no reason why this should converge without using additional
properties of $\alpha$. In fact, we will not be able to show that the limit
as $\epsilon \to 0$ of $h_\epsilon(x)$ exists, but we will nevertheless 
construct a smooth solution to eq.~(\ref{psh}).

To proceed, we rewrite eq.~(\ref{hedef}) as
\ben
h_\epsilon(x) = - \int_0^\infty {\rm e}^{-\epsilon \tau} 
\hat{\phi}^*_\tau\alpha(x) \, \dd\tau \, + \frac{\kappa}{\epsilon} \,,
\een
where $\hat{\phi}^*_\tau$ denotes the pull-back map on tensor fields
associated with $\hat{\phi}_\tau$. Taking the gradient of this equation
and using eq.~(\ref{dalpha2}), we obtain 
\ben
\dd h_\epsilon(x) = - \int_0^\infty {\rm e}^{-\epsilon \tau} 
\hat{\phi}^*_\tau (\dd \alpha)(x) \, \dd\tau 
= \frac{1}{2} \int_0^\infty {\rm e}^{-\epsilon \tau} 
\hat{\phi}^*_\tau (\pounds_s \beta)(x) \, \dd\tau \,, 
\een
where here and in the following we
use differential forms notation and omit 
tensor indices. Since $\pounds_s$ clearly
commutes with $\hat{\phi}^*_\tau$ and since $\pounds_s$ is just the
derivative along the orbit over which we are integrating, we can
integrate by parts to obtain
\ben\label{ddh}
\dd h_\epsilon(x) = - \half \beta(x) + 
\frac{\epsilon}{2} \int_0^\infty \e^{-\epsilon \tau}
\, \hat{\phi}_\tau^* \beta(x) \, \dd \tau \,.   
\een

It follows from the von Neumann ergodic theorem\footnote{ 
Here, the theorem is applied to the case of a 
tensor field $T$ of type $(k,l)$
on a compact Riemannian manifold $\Sigma$, rather than a scalar function, and 
where the measure preserving map is a smooth one-parameter family of 
isometries acting on $T$ via the pull back. To prove this generalization, we
note that a tensor field $T$ of type $(k,l)$ on a manifold $\Sigma$
may be viewed as a function on the fiber bundle, $B$, of 
all tensors of type $(l,k)$
over $\Sigma$ that satisfies the additional property that this function is 
linear on each fiber. Equivalently, we may view $T$ as a function, $F$, on the
bundle, $B'$, of unit norm tensors of type $(l,k)$ that satisfies a
corresponding linearity property. A Riemannian metric on $\Sigma$ 
naturally gives rise to a Riemannian metric (and, in particular, a
volume element) on $B'$, and $B'$ is compact provided that $\Sigma$ is 
compact. Since the isometry flow on $\Sigma$ naturally induces a volume
preserving flow on $B'$, we may apply the von Neumann ergodic theorem
to $F$ to obtain the orbit averaged function $F^*$. Since $F^*$ will satisfy 
the appropriate linearity property on each fiber, we thereby obtain
the desired orbit averaged tensor field $T^*$.} 
(see eq.~\eqref{vnet}) that the limit 
\ben
\label{lim1}
\lim_{\epsilon \to 0} \epsilon 
\int_0^\infty \e^{-\epsilon \tau}
 \, \hat{\phi}_\tau^* \beta(x)  \, \dd \tau  = \beta^*(x) \, ,   
\een
exists in the sense of $L^p(\Sigma)$. 
Furthermore, the limit in the sense of $L^p(\Sigma)$ 
also exists of all $x$-derivatives of the left side. Indeed, 
because $\hat{\phi}_\tau$ is an isometry commuting with 
the derivative operator $D_a$ 
of the metric $\gamma_{ab}$, we have 
\ben
D_{c_1} \cdots D_{c_k} \left(
\epsilon 
\int_0^\infty \e^{-\epsilon \tau}
 \, \hat{\phi}_\tau^* \beta_a(x)  \, \dd \tau  \right) 
=
\epsilon 
\int_0^\infty \e^{-\epsilon \tau}
 \, \hat{\phi}_\tau^* D_{c_1} \cdots D_{c_k} \beta_a(x)  \, \dd \tau
 \, .
\een
The expression on the right side converges in $L^p$, as $\epsilon \to
0$ by the von Neumann ergodic theorem, meaning that 
\ben
\dd h_\epsilon \to - \half (\beta - \beta^*) \quad 
\text{ in $W^{k,p}(\Sigma)$ as $\epsilon \to 0$} \,, 
\een
for all $k \ge 0,\, p \ge 1$, where  $W^{k,p}(\Sigma)$ denotes
the Sobolev space of order $(k,p)$. By the 
Sobolev embedding theorem, 
\ben
 C^m(\Sigma) \hookleftarrow W^{k,p}(\Sigma) \quad \text{for $k>m+(n-2)/p$}\,, 
\een
where the embedding is continuous with respect to the sup norm 
on the all derivatives in the space $C^m$, i.e., 
$\sup_\Sigma |D^m \psi(x)| \le {\rm const.} \|\psi\|_{W^{k,p}}$ for 
all $\psi \in C^m$. Thus, convergence 
of the limit~\eqref{lim1} actually occurs in the sup norms on $C^m$. 
Thus, in particular, 
$\beta^* \in C^\infty = \cap_{m\ge 0} C^m$.  

Now pick an arbitrary $x_0 \in \Sigma$, and define $F_\epsilon$ by
\ben
h_\epsilon(x) - h_\epsilon(x_0) = \int_{C(x)} \dd h_\epsilon 
                                = F_\epsilon(x) \,,
\een
where the integral is over any smooth path $C(x)$ connecting $x_0$ and $x$. 
This integral manifestly does not depend upon the choice of $C(x)$,
independently of the topology of $\Sigma$. 
By what we have said above, the function $F_\epsilon$ is smooth, with 
a smooth limit 
\ben
 F(x) = \lim_{\epsilon \to 0} F_\epsilon(x) 
      = - \half\int_{C(x)} (\beta - \beta^*)\,, 
\een
which is independent of the choice of $C(x)$. Furthermore, the convergence
of $F_\epsilon$ and its derivatives to $F$ and its derivatives is uniform.
Now, by inspection, $F_\epsilon$ is a solution 
to the differential equation 
\ben
\pounds_s F_\epsilon(x) = \alpha(x) - \kappa - \epsilon F_\epsilon(x) - 
\epsilon h_\epsilon(x_0) \, .
\een
Furthermore, the limit
\bena
\lim_{\epsilon \to 0} \epsilon h_\epsilon(x_0) 
 &=& - \lim_{\epsilon \to 0} \epsilon \int_0^\infty 
       \e^{-\tau \epsilon} 
       \left[\alpha(\hat{\phi}_\tau(x_0)) - \kappa\right] \, \dd \tau 
\non \\
&=& \kappa - \lim_{S \to \infty} \frac{1}{S} \int_0^S 
                                 \alpha(\hat{\phi}_\tau(x_0))\,\dd \tau 
 = 0 
\eena
exists by the ergodic theorem, and vanishes 
by Lemma~1. Thus, the smooth, limiting quantity  
$F = \lim_{\epsilon \to 0} F_\epsilon$ satisfies the desired differential 
equation~\eqref{psh}. 
$\Box$ \hfill  

\bigskip

We now define a new set of Gaussian null coordinates $(\tilde{u}, 
\tilde{r}, \tilde{x}^A)$
as follows. Define $\tilde{u}$ on $\Sigma$ to be a smooth solution to
eq.~(\ref{tildeu}), whose existence is guaranteed by Lemma~2. 
Extend $\tilde{u}$ to $H$ by eq.~(\ref{normali}). 
It is not difficult to verify that $\tilde{u}$ is given explicitly by 
\ben
\label{tilu}
\tilde u(x) 
 = \int_0^{u(x)} \left[ 
                       f\left(\pi \circ\phi_{-\tau}(x) \right) 
                 \right]^{-1}\dd \tau 
  + \frac{1}{\kappa} \ln f(\pi(x)) - \frac{1}{\kappa} h(\pi(x)) \,, 
\een
where $f$ and $h$ are smooth solutions to 
eqs.~(\ref{f}) and (\ref{psh}), respectively, on
$\Sigma$ and
\ben
\pi: H \to \Sigma, \quad x \mapsto \pi(x) 
\een
is the map projecting any point $x$ in $H$ to the point $\pi(x)$ on 
the cross section $\Sigma$ on the null generator through $x$. Let
$\tilde{\Sigma}$ denote the surface $\tilde{u} = 0$ on $H$. Then our desired
Gaussian null coordinates $(\tilde{u}, \tilde{r}, \tilde{x}^A)$ are the
Gaussian null coordinates associated with $\tilde{\Sigma}$.
The corresponding fields 
${\tilde \alpha}, {\tilde \beta}_a, 
{\tilde \gamma}_{ab}$ satisfy all of the properties derived above for 
$\alpha, \beta_a, \gamma_{ab}$ and, in addition, satisfy the condition
that $\tilde{\alpha} = \kappa$ is constant on $H$.

Now let $K^a = \tilde{n}^a$. We have previously shown that
$\pounds_{\tilde{n}} \tilde \gamma_{ab}=0$ on $H$, since this relation holds
for any choice of Gaussian null coordinates. However, since our new
coordinates have the property that 
$\tilde \alpha = \kappa$ is constant on $H$, we clearly have that 
$\pounds_{\tilde{n}} \tilde \alpha = 0$ on $H$. 
Furthermore, for our new coordinates,
eq.~(\ref{dalpha}) immediately yields 
$\pounds_{\tilde{n}} \tilde \beta_a = 0$ on $H$.
Thus, we have proven that all of the relations in eq.~(\ref{cid}) hold
for $m=0$.

We next prove that the 
equation $\pounds_{\tilde \ell} \, \pounds_{\tilde n}  \tilde
\gamma_{ab} = 0$ holds on $H$. Using what 
we already know about 
$\tilde \beta_a, \tilde \gamma_{ab}$ 
and taking the Lie-derivative $\pounds_{\tilde n}$ of the Einstein equation 
$R_{ab} (\partial/\partial \tilde{x}^A)^a (\partial/\partial \tilde{x}^B)^b=0$ 
(see eq.~(\ref{ABR}) of Appendix~\ref{sect:A}), we get 
\ben
0 = \pounds_{\tilde n}
    \left[ \pounds_{\tilde n} \pounds_{\tilde \ell} 
           \tilde \gamma_{ab} 
          + {\kappa} \, \pounds_{\tilde \ell}\tilde \gamma_{ab}
    \right] \,,  
\label{eq:nlgamma}
\een
on $H$.
Since $t^a = \tilde n^a + \tilde s^a$, with $\tilde s^a$ tangent 
to $\tilde \Sigma(\tilde{u})$, and since all quantities appearing
in eq.~(\ref{eq:nlgamma}) are Lie derived by $t^a$, we may replace 
in this equation all Lie derivatives 
$\pounds_{\tilde n}$ by $- \pounds_{\tilde s}$. 
Hence, we obtain
\ben
\label{31}
0 = \pounds_{\tilde s} \left[ \pounds_{\tilde s} \pounds_{\tilde \ell}
  \tilde \gamma_{ab}
 - {\kappa} \pounds_{\tilde \ell} \tilde \gamma_{ab} \right] \,,  
\een
on $\tilde \Sigma$. 
Now, write $L_{ab} = \pounds_{\tilde \ell} \tilde \gamma_{ab}$. We
fix $x_0 \in \tilde{\Sigma}$ and view eq.~(\ref{31}) as an 
equation holding at $x_0$ for the pullback, $\hat{\phi}^*_\tau L_{ab}$, 
of $L_{ab}$ to $x_0$, where
$\hat{\phi}_\tau: \tilde \Sigma \to \tilde \Sigma$ now 
denotes the flow of $\tilde s^a$. Then eq.~(\ref{31}) can be rewritten as
\ben
\frac{\dd}{\dd \tau} \left[\e^{\kappa \tau} \frac{\dd}{\dd \tau}
(\e^{-\kappa \tau} \hat{\phi}^*_\tau L_{ab})\right] = 0 \,. 
\een
Integration of this equation yields
\ben
\e^{\kappa \tau} \frac{\dd}{\dd \tau}
(\e^{-\kappa \tau}  \hat{\phi}^*_\tau L_{ab}) = - \kappa C_{ab} \,, 
\een
where $C_{ab}$ is a tensor at $x_0$ that
is independent of $\tau$. Integrating this equation (and
absorbing constant factors into $C_{ab}$), we obtain
\ben
\label{i}
{\hat{\phi}}_\tau^* \, L_{ab}      
- \e^{\kappa \tau} L_{ab}
= (1-\e^{\kappa \tau}) C_{ab} \,.  
\een
However, since $\hat{\phi}_\tau$ is a Riemannian isometry, 
each orthonormal frame component of 
${\hat{\phi}}_\tau^* \, L_{ab}$ at $x_0$ is uniformly bounded 
in $\tau$ by the Riemannian norm of $L_{ab}$, i.e.,
$\sup\{(L^{ab} L_{ab}(x))^{1/2}; \,\, x \in \tilde \Sigma\}$. 
Consequently, the limit of eq.~(\ref{i}) as 
$\tau \rightarrow \infty$ immediately yields 
\ben
L_{ab} = C_{ab}
\een
from which it then immediately follows that
\ben
{\hat{\phi}}_\tau^* \, L_{ab} = L_{ab} \, .
\een
Thus, we have
$\pounds_{\tilde s} \pounds_{\tilde \ell} \tilde \gamma_{ab}= 0$, 
and therefore    
$\pounds_{\tilde n} \pounds_{\tilde \ell} \tilde \gamma_{ab}
=\pounds_{\tilde \ell} \, \pounds_{\tilde n}  \tilde
\gamma_{ab} = 0$ on $H$,
as we desired to show.

Thus, we now have shown that the first equation in~\eqref{cid} 
holds for $m=0,1$, and that the other equations hold for $m=0$, 
for the tensor fields associated with the ``tilde" 
Gaussian null coordinate system, and $K = \tilde n$.
In order to prove that eq.~(\ref{cid}) holds for all $m$, we proceed
inductively. Let $M \ge 1$, and assume inductively that 
the first of equations~\eqref{cid} holds for all $m \le M$, and that the 
remaining equations hold for all $m \le M-1$. Our task is 
to prove that these statements then also hold when $M$ is
replaced by $M+1$. 
To show this, we apply the operator $(\pounds_{\tilde \ell})^{M-1}
\pounds_{\tilde n}$
to the Einstein equation $R_{ab} \tilde n^a
\tilde \ell^b=0$ (see eq.~(\ref{nlR})) and restrict to $H$. 
Using the inductive hypothesis, 
one sees that 
$(\pounds_{\tilde \ell})^M (\pounds_{\tilde n} \tilde \alpha)=0$ on $H$,
thus establishes the second equation in~\eqref{cid} for $m \le M$. 
%
Next, we apply the operator $(\pounds_{\tilde \ell})^{M-1}
\pounds_{\tilde n}$ to the Einstein equation $R_{ab}
(\partial/\partial \tilde x^A)^a
\tilde \ell^b=0$ (see eq.~(\ref{lAR})), 
and restrict to $H$. Using the inductive
hypothesis, one sees that 
$(\pounds_{\tilde \ell})^M (\pounds_{\tilde n} \tilde \beta_a)=0$ 
on $H$, thus establishes the third equation in~\eqref{cid} for $m \le M$. 
%
%
Next, we apply the operator $(\pounds_{\tilde \ell})^M \pounds_{\tilde n}$ 
to the Einstein equation $R_{ab} 
(\partial/\partial \tilde x^A)^a(\partial/\partial \tilde x^B)^b
=0$ (see eq.~(\ref{ABR})), and restrict to $H$. Using the inductive 
hypothesis and the above results 
$(\pounds_{\tilde \ell})^M (\pounds_{\tilde n} \tilde \alpha)=0$ 
and $(\pounds_{\tilde \ell})^M (\pounds_{\tilde n} \tilde \beta_a)=0$, 
one sees that the tensor field 
$L_{ab}^{(M+1)} \equiv (\pounds_{\tilde \ell})^{M+1} \tilde \gamma_{ab}$ 
satisfies a differential equation of the form 
\ben
\pounds_{\tilde n}[\pounds_{\tilde n} L_{ab}^{(M+1)} 
                   + (M+1)\kappa \,L_{ab}^{(M+1)}] = 0 
\een
on $H$. By the same argument as given above for $L_{ab}$, 
it follows that $\pounds_{\tilde n} \, L_{ab}^{(M+1)} = 0$. 
This establishes the first equation 
in~\eqref{cid} for $m \le M+1$, and closes the induction loop.

\medskip

Thus far, we have assumed only that the spacetime metric is 
smooth ($C^\infty$). However, if we now assume that the spacetime is
real analytic, and that $H$ is an analytic submanifold, then it can be
shown that the vector field $K^a$ that we have defined above is, 
in fact, analytic. To see this, first note that if the cross section
$\Sigma$ of $H$ is chosen to be analytic, then our Gaussian null coordinates 
are analytic, and, consequently, so is any quantity defined in terms of them, 
such as $n^a$ and $\ell^a$. Above, $K^a$ was defined in terms of a certain 
special Gaussian normal coordinate system that was obtained from 
a geometrically special cross section. That cross 
section was obtained by a change~\eqref{tilu} 
of the coordinate $u$. Thus, to show 
that $K^a$ is analytic, we must show that this change of coordinates
is analytic. By eq.~\eqref{tilu}, this will be the case provided that $f$
and $h$ are analytic. We prove this in Appendix~\ref{sect:C}. 

Since $g_{ab}$ and $K^a$ are analytic, so is $\pounds_K g_{ab}$.
It follows immediately from the fact that this quantity and all of its
derivatives vanish at any point of $H$ that $\pounds_K g_{ab} = 0$ where
defined, i.e., within the region where the Gaussian null coordinates
$(\tilde{u}, \tilde{r}, \tilde{x}^A)$ are defined. This proves existence
of a Killing field $K^a$ in a neighborhood of the horizon. We
may then extend $K^a$ by analytic continuation.
Now, analytic continuation
need not, in general, give rise to a single-valued extension, 
so we cannot conclude that there exists a Killing field
on the entire spacetime. However, 
by a theorem of Nomizu~\cite{Nomizu60} (see also~\cite{Chr97}),
if the underlying domain is simply connected,
then analytic continuation does give rise to a single-valued extension.
By the topological censorship theorem~\cite{Galloway99,Galloway01}, 
the domain of outer
communication has this property. Consequently, there exists
a unique, single valued extension of $K^a$ to the 
domain of outer communication, i.e., the exterior of the black hole 
(with respect to a given end of infinity). Thus, in the analytic case, 
we have proven the following theorem:

\paragraph{Theorem 1:} Let $(M,g_{ab})$ 
be an analytic,
asymptotically flat $n$-dimensional solution of the vacuum Einstein equations
containing a black hole and possessing a
Killing field $t^a$ with complete orbits which are timelike near infinity. 
Assume that a connected component, $H$, of the
event horizon of the black hole 
is analytic and is topologically $\mr \times \Sigma$, 
with $\Sigma$ compact 
and that $\kappa \neq 0$ (where $\kappa$ is defined 
eq.~\eqref{kdef} above). Then there exists a Killing field $K^a$, defined
in a region that covers $H$ and the entire domain of outer communication,
such that $K^a$ is normal to the horizon and $K^a$ commutes with $t^a$. 

\bigskip 

The assumption of analyticity in this theorem can be partially removed 
in the following manner, using an argument
similar to that given in~\cite{FRW99}. Since $\kappa > 0$, the arguments
of~\cite{RW96} show that the spacetime can be extended, if necessary,
so that $H$ is a proper subset of a regular bifurcate null surface $H^*$ 
in some enlarged spacetime $(M^*,g^*)$. 
We may then consider the characteristic initial value formulation for 
Einstein's equations~\cite{Rendall90,MzH90,Friedrich91} on this 
bifurcate null surface. Since the extended spacetime is smooth, the initial
data induced on this bifurcate null surface should be regular. Since
this initial data is invariant under the
orbits of $K^a$, it follows that the solution to which this data gives
rise will be invariant under a corresponding one-parameter group of
diffeomorphisms in the domain of dependence, $D(H^*)$,
of $H^*$. Thus, if one merely assumes that the spacetime is smooth, existence
of a Killing field in $D(H^*)$ holds. However, since $D(H^*)$ lies inside
the black hole, this argument does not show existence of a Killing field
in the domain of outer communications. Interestingly, if one assumes that
the spacetime is analytic---so that existence of a Killing field in the
domain of outer communications follows from the above analytic continuation
arguments---then this argument shows that the Killing field known to 
exist in the domain of outer communications also can
be extended to all of $D(H^*)$.

\section{Proof of existence of rotational Killing fields} \label{sect:3}

We proved in the previous section that if 
the quantity $\kappa$ defined by eq.~\eqref{kdef} is non-vanishing, then 
there exists a vector field $K^a$ in a neighborhood of $H$ which is
normal to $H$ and is such that the 
equations~\eqref{civ1} hold. As explained at the end of the previous section,
in the analytic case, this implies the existence of a Killing field
normal to the horizon in a region containing the horizon and the domain
of outer communication. Since we are considering the case where 
$t^a$ is not pointing along the null
generators of $H$, the Killing field $K^a$ is distinct from $t^a$. 
Hence, their
difference $S^a \equiv \tilde{s}^a =t^a-K^a$ 
is also a nontrivial Killing field. 
There are two cases to consider: 
\begin{enumerate}
\item
The Killing field $S^a$ has closed orbits, or
\item 
The Killing field $S^a$ does not have closed orbits. 
\end{enumerate}
Only the first case can occur in $4$-dimensions. In the first case, it follows
immediately that the Killing field $S^a$ corresponds to a rotation
at infinity. The purpose of this section is to show that, 
in the second case, even though the orbits of $S^a$ are not closed, there
must exist 
$N \geq 2$ mutually 
commuting
Killing fields, $\varphi_{(1)}^a, \dots, \varphi_{(N)}^a$, 
which possess closed orbits with period $2 \pi$ and are such 
that 
\ben
S^a = \Omega_1^{} \varphi_{(1)}^a + \dots +  \Omega_N^{} \varphi_{(N)}^a \, , 
\label{Sphi}
\een
for some constants $\Omega_i$, all of whose ratios are irrational.

To simplify notation, throughout this section, we omit the ``tildes''
on all quantities, i.e., in this section 
$\ell^a, n^a, \Sigma, u, r, \alpha, \beta_a, \gamma_{ab}$ 
denote the quantities associated with our preferred Gaussian null coordinates.
The Killing field $S^a$ satisfies a
number of properties that follow immediately from the construction of
$K^a=n^a$ given in the previous section.
First, since $\pounds_K r = 0 =
\pounds_t r$ and since $\pounds_K \, \ell^a = 0 = \pounds_t \, \ell^a$,
it follows that $S^a$ also
satisfies these properties, i.e., $\pounds_S r = 0$ and 
$\pounds_S \,\ell^a = 0$. Similarly, since $\pounds_K u = 1 = \pounds_t u$, 
we also have $\pounds_S u = 0$. 
In addition, since $K^a$ commutes with $t^a$, so does $S^a$.
Thus, $S^a$ is tangent to the surfaces of constant $(u,r)$, and 
commutes with $\ell^a$ and $t^a$. 
Finally, it follows immediately from eq.~\eqref{cid} and 
eqs.~(\ref{ta})-(\ref{tc}) that $S^a$ 
also satisfies the analog of eq.~\eqref{cid}. 

To proceed, we focus attention now on the Riemannian manifold
$(\Sigma, \gamma_{ab})$ and make arguments similar to those given 
in \cite{IM92}. 
Let $\mathcal G$ denote the isometry group of
$(\Sigma, \gamma_{ab})$. Then $\mathcal G$ is a compact Lie group.
Let ${\mathcal H} \subset {\mathcal G}$ denote the one-parameter
subgroup of $\mathcal G$ generated by the Killing field $S^a$ on
$(\Sigma, \gamma_{ab})$. Then the closure, $\overline{\mathcal H}$, of
${\mathcal H}$ is a closed subgroup of ${\mathcal G}$, and hence is a
Lie subgroup. Since $\mathcal H$ is abelian, so is $\overline{\mathcal H}$,
and since $\mathcal G$ is compact, it follows that $\overline{\mathcal H}$
is a torus. Let $N = {\rm dim} (\overline{\mathcal H})$. Since the
$N$-dimensional torus can be written as the direct product of $N$
factors of $U(1)$, it follows that the isometries in $\overline{\mathcal
H}$ are generated by $N$ commuting Killing vector fields,
$\varphi_{(1)}^a, \dots, \varphi_{(N)}^a$, which possess closed orbits
on $\Sigma$ with period $2 \pi$. Since the isometry subgroup $\mathcal
H$ generated by $S^a$ is dense in $\overline{\mathcal H}$, it follows that,
on $\Sigma$, $S^a$ must be a linear combination of these Killing vector
fields of the form (\ref{Sphi}).

Since, as we have noted above, $S^a$ 
satisfies the analog of eq.~\eqref{cid},
the diffeomorphisms on $\Sigma$ corresponding to every element of
$\mathcal H$ leave invariant each tensor field on $\Sigma$ of the form
\ben
T_{(k)} = 
\begin{cases}
\underbrace{\pounds_\ell \, \cdots \, \pounds_\ell}_{m+1 \,\, {\rm times}} 
\gamma_{ab} \\
\underbrace{\pounds_\ell \, \cdots \, \pounds_\ell}_{m \,\, {\rm times}} 
\beta_a \\
\underbrace{\pounds_\ell \, \cdots \, \pounds_\ell}_{m \,\, {\rm times}} 
\, \alpha \, , 
\end{cases}
\een
for all $m\ge 0$. Since $\mathcal
H$ is dense in $\overline{\mathcal H}$, each $T_{(k)}$ also must be
invariant under the diffeomorphisms corresponding to the elements
of $\overline{\mathcal H}$.
Consequently, the Killing fields
$\varphi^a_{(1)}, \dots, \varphi^a_{(N)}$ 
Lie derive all $T_{(k)}$ on $\Sigma$.
We now extend each $\varphi^a_{(j)}$ to a vector field defined in an
entire neighborhood of $H$ as follows. 
First, we Lie-drag $\varphi^a_{(j)}$ from $\Sigma$ to $H$ via the vector
field $K^a=n^a$. Then we Lie-drag the resulting vector field defined
on $H$ off the horizon via the vector field $\ell^a$. The vector field
(denoted again by $\varphi^a_{(j)}$), which has now been defined in an entire
neighborhood of $H$, satisfies the following properties throughout this
neighborhood: (i) $\varphi^a_{(j)}$
commutes with both $n^a$ and $\ell^a$. (ii) $\varphi^a_{(j)}$ 
satisfies the analog of eq.~\eqref{cid}. Property (ii) implies
that in the analytic case, $\varphi^a_{(j)}$ is a Killing field
of the spacetime metric. As was the 
case for $K^a$, we may then uniquely extend $\varphi^a_{(j)}$ 
as a Killing field to the entire
domain of outer communication. That this extended Killing field (which we also
denote by $\varphi^a_{(j)}$) must have closed orbits can be seen as follows:
The orbits of $\varphi^a_{(j)}$ on $\Sigma$ are closed with period $2 \pi$. 
Thus, if we consider the flow of $\varphi^a_{(j)}$ by 
parameter $2 \pi$, any point
$x \in \Sigma$ will be mapped into itself, and vectors at $x$ that are 
tangent to $\Sigma$ also will get mapped into themselves.
Furthermore, since
$\varphi^a_{(j)}$
commutes with $n^a$ and $\ell^a$ tangent vectors at $x$ that are orthogonal
to $\Sigma$ also will get mapped into themselves. Consequently, the isometry
on the spacetime corresponding to the action of $\varphi^a_{(j)}$ 
by parameter $2 \pi$
maps point $x$ into itself and maps each vector at $x$ into itself. 
Consequently, this isometry is the identity map in any connected region where
it is defined.
Thus, we have shown:

\paragraph{Theorem 2:} Let $(M,g_{ab})$ 
be an analytic,
asymptotically flat $n$-dimensional solution of the vacuum Einstein equations
containing a black hole and possessing a
Killing field $t^a$ with complete orbits which are timelike near infinity. 
Assume that a connected component, $H$, of the
event horizon of the black hole 
is analytic and is topologically $\mr \times \Sigma$, 
with $\Sigma$ compact 
and that $\kappa \neq 0$ (where $\kappa$ is defined 
eq.~\eqref{kdef} above). If $t^a$ is not tangent to the 
generators of $H$, then there exist
mutually commuting Killing fields 
$\varphi^a_{(1)}, \dots, \varphi^a_{(N)}$ (where $N\ge 1$) with 
closed orbits with period $2 \pi$ which are defined 
in a region that covers $H$ and the entire domain of outer communication.
Each of these Killing fields commute with $t^a$, and $t^a$ can be written as
\ben
  t^a = K^a + \Omega_1^{} \varphi_{(1)}^a + \dots 
            +  \Omega_N^{} \varphi_{(N)}^a \, , 
\een
for some constants $\Omega_i$, all of whose ratios are irrational, where
$K^a$ is the horizon Killing field whose existence is guaranteed by
Theorem~1.

\bigskip

Theorem 2 shows that the null geodesic generators of the event horizon
rotate rigidly with respect to infinity.

\paragraph{Remarks:} 

\noindent

1) As in the case of $K^a$, in the non-analytic
case the Killing fields 
$\varphi^a_{(1)}, \dots, \varphi^a_{(N)}$ can be proven to exist in $D(H^*)$
(see the end of section~\ref{sect:2}).

2) If the orbits of $S^a$ are closed on $\Sigma$---i.e., equivalently,
if the orbits of $t^a$ map each generator of $H$ to itself after some
period $P$---then the above 
argument shows that $S^a$ itself is a Killing field with closed orbits.
As previously noted in the introduction, in the case of 
$4$-dimensional spacetimes, the orbits of $S^a$ on 
$\Sigma$ are always closed. However, the orbits of $S^a$ on 
$\Sigma$ need not be closed when $n>4$. 
For example, on the round $3$-sphere $S^3$, 
one can take an incommensurable linear combination 
of two commuting Killing fields with closed orbits to 
obtain a Killing field with non-closed orbits. 
This possibility is realized for $S^a$ in suitably chosen 
$5$-dimensional Myers-Perry black hole solutions~\cite{MP86}. Our theorem shows
that if the orbits of $S^a$ fail to be closed, then the spacetime must
admit at least two linearly independent rotational Killing fields.

3) In $4$-dimensions,
if $t^a$ is normal to the horizon, then Thm.~3.4 of
\cite{SW92} (applied to the vacuum or Einstein-Maxwell cases) 
shows that the exterior
region must be static. 
The proof of this result makes use of the fact
that there exists a bifurcation surface (i.e.,
that we are in the non-degenerate case $\kappa \neq 0$), and that there
exists a suitable foliation of the exterior region by 
maximal surfaces; the existence of such a foliation was 
proven in~\cite{CW94}. The arguments of \cite{SW92} 
generalize straightforwardly to higher dimensions. Thus, the
staticity of higher dimensional vacuum or Einstein-Maxwell stationary
black holes with $t^a$ normal to the horizon must hold provided that
the arguments of~\cite{CW94} also generalize suitably to 
higher dimensions. 
\footnote{ 
It has recently been shown in \cite{Rogatko05} that  
this is indeed the case. 
 } 
It should be noted that $n$-dimensional static vacuum
(and Maxwell-dilaton) black hole spacetimes with a standard null 
infinity of topology ${\mathscr  I} \cong S^{n-2} \times \mr$ 
were shown to be essentially unique by 
Gibbons et al.~\cite{Gibbons02a}, and are, in particular, 
spherically symmetric. 
However, static spacetimes with a non-trivial topology at infinity 
do not have to have any extra Killing fields~\cite{Gibbons02b}. 
\footnote{ 
See also~\cite{Rogatko02,Rogatko03,Rogatko04a} for uniqueness 
results of higher dimensional static black holes 
and~\cite{Morisawa:04,Rogatko04b} for uniqueness results of 
some restricted class of $5$-dimensional stationary black holes.  
 } 

\section{Matter fields} \label{sect:4}

The analysis of the foregoing sections can be generalized to include various 
matter sources, and we now illustrate this by discussing several examples. 
The Einstein equation with matter is 
\ben
R_{ab} = T_{ab} - \frac{1}{n-2} g_{ab} T^c{}_c \,. 
\een
The simplest matter source is a cosmological constant 
$T_{ab} = - \Lambda g_{ab}$. The only significant change resulting 
from the presence of a cosmological constant is a change in the asymptotic 
properties of the spacetime. For the case of a negative cosmological constant,
we can consider spacetimes that are asymptotically AdS rather than 
asymptotically flat. For asymptotically AdS spacetimes, 
$\mathscr I$ is no longer 
null, but is instead timelike. However, the only place where we
used the character of $\mathscr I$ in our arguments was to conclude
that the Killing field $t^a$ is nowhere vanishing on $H$,
and therefore generates a suitable foliation $\Sigma(u)$ of $H$ by
cross sections.  This argument goes through without change in the
asymptotically AdS case, as do all subsequent arguments in our proof. Thus,
the rigidity theorem holds without modification in the case of a negative
cosmological constant. For a positive
cosmological constant, 
$\mathscr I$ would have a spacelike character
and it is not clear precisely what should be assumed about the behavior
of $t^a$ near $\mathscr I$. Nevertheless, our results apply straightforwardly
to any ``horizon'' that is the boundary of the past of any complete, timelike
orbit of $t^a$. In this sense, our rigidity theorem holds for both
black hole event horizons and cosmological horizons.

For Maxwell fields the field equations and stress tensor are 
\ben
\nabla^a F_{ab} = 0 \,, \quad \nabla_{[a} F_{bc]} = 0\,, \quad 
T_{ab} = F_{ac} F_b{}^c
               - \frac{1}{4} g_{ab} F^{cd} F_{cd} \, .
\een
We assume that both metric and Maxwell tensor are invariant under $t^a$, 
i.e., $\pounds_t g_{ab} = 0 = \pounds_t F_{ab}$. 
In parallel with the vacuum case, we wish to show that 
there exists a vector field $K^a$ tangent to the generators of the horizon 
satisfying
\ben
\label{cid3}
\underbrace{\pounds_\ell \, \pounds_\ell \, \cdots \,
\pounds_\ell}_{m \,\, {\rm times}} 
(\pounds_K g_{ab}) = 0 \,, \quad 
\underbrace{\pounds_\ell \, \pounds_\ell \, \cdots \,
\pounds_\ell}_{m \,\, {\rm times}} 
(\pounds_K F_{ab}) = 0 \,, \quad 
m=0,1,2, \dots, 
\een
on $H$.
To analyze these equations we introduce a Gaussian null coordinate system as 
above, and correspondingly decompose the field strength tensor as 
\ben
\label{decomp}
F_{\mu\nu} \, \dd x^\mu \wedge \dd x^\nu 
= S \, \dd u \wedge \dd r 
 + V_A \, \dd u \wedge \dd x^A 
 + W_A \, \dd r \wedge \dd x^A 
 + U_{AB} \, \dd x^A \wedge \dd x^B \, .
\een
We write 
\ben
V_a = V_A (\dd x^A)_a \,, \quad W_a = W_A(\dd x^A)_a \,, \quad 
U_{ab} = U_{AB}(\dd x^A)_a (\dd x^B)_b \,.  
\label{VWU}
\een
It follows from $\pounds_t F_{ab} = 0$ that $\pounds_t S = \pounds_t V_a 
= \pounds_t W_a = \pounds_t U_{ab} = 0$. 
As in the vacuum case, we take the candidate Killing field 
$K^a$ to be $n^a$, where $n^a$ is
the vector field associated with a suitable Gaussian null coordinate
system to be determined.
Equations~\eqref{cid3} are then 
equivalent to eqs.~\eqref{cid} together with 
\bena
\label{cid4}
\underbrace{\pounds_\ell \, \pounds_\ell \, \cdots \,
\pounds_\ell}_{m \,\, {\rm times}} 
\left(\pounds_n S  \right) &=& 0 \, ,
\non\\
\underbrace{\pounds_\ell \, \pounds_\ell \, \cdots \,
\pounds_\ell}_{m \,\, {\rm times}} 
\left(\pounds_n V_a \right) &=& 0 \, ,
\non\\
\underbrace{\pounds_\ell \, \pounds_\ell \, \cdots \,
\pounds_\ell}_{m \,\, {\rm times}} 
\left(\pounds_n W_a \right) &=& 0 \, , 
\non\\ 
\underbrace{\pounds_\ell \, \pounds_\ell \, \cdots \,
\pounds_\ell}_{m \,\, {\rm times}} 
\left(\pounds_n U_{ab} \right) &=& 0 \,.  
\eena
The Maxwell equations, the Bianchi identities, and the stress tensor 
are presented in Appendix~\ref{sect:B}. The Raychaudhuri equation now 
gives $\pounds_n \gamma_{ab} = 0 = V_a$ on 
$H$. Then, the Bianchi identity \eqref{BI:uAB} 
yields $\pounds_n U_{ab} = 0$. The Maxwell equation~\eqref{maxwell:r} 
yields $\pounds_n S = 0$ on $H$. 
Furthermore, we have $T_{ab} n^a (\partial/\partial x^A)^b = 0$ 
(see eq.~\eqref{T:uA}), from which it follows in view of Einstein's equation 
that also $R_{ab} n^a (\partial/\partial x^A)^b = 0$ on $H$. 
We may now argue in precisely the same way as in the vacuum case that, 
by a suitable choice of Gaussian null coordinates, we can achieve 
that $\pounds_n \alpha = 0 = \pounds_n\beta_a$ on $H$. 
It then follows from eqs.~\eqref{T:AB} and \eqref{T:trace} that 
$\pounds_n [T_{ab} (\partial/\partial x^A)^a (\partial/\partial x^B)^b] = 0$ 
and $\pounds_n {T^a}_a = 0$ on $H$, 
which in view of Einstein's equation means that 
$\pounds_n [R_{ab}(\partial/\partial x^A)^a (\partial/\partial x^B)^b] = 0$ 
on $H$. This may in turn be used to argue, 
precisely as in the vacuum case, that 
$\pounds_\ell \pounds_n \gamma_{ab} = 0$ on $H$. 
Taking a Lie derivative $\pounds_n$ of the Maxwell equation \eqref{maxwell:A}  
and the Bianchi identity \eqref{BI:urA} then leads to the equation 
\ben
\pounds_n \left[ \pounds_n W_a + \kappa \, W_a \right] = 0 \,, 
\een
on $H$. Since $W_a$ is Lie derived by $t^a$ and since 
$t^a = n^a+s^a$ as in the vacuum case, this 
equation may alternatively be written as 
\ben
\pounds_s \left[ \pounds_s W_a - \kappa \, W_a \right] = 0 \,. 
\een
Integration gives 
\ben
\label{above}
\hat \phi_\tau^* W_a - \e^{\kappa \tau} W_a = (1-\e^{\kappa \tau}) C_a \, , 
\een
where $C_a$ is a 1-form field on $\Sigma$ independent of $\tau$, and where 
$\hat \phi_\tau$ is again the flow of $s^a$. The same type of 
argument following eq.~\eqref{i} then implies that $\pounds_n W_a = 0$
on $H$. We have thus shown that 
all eqs.~\eqref{cid4} and~\eqref{cid} for $m=0$ and the first equation 
in~\eqref{cid} for $m=1$ are satisfied on the horizon. 
The remainder of the argument closely parallels the vacuum case. 

\bigskip 
\begin{center}
{\bf Acknowledgements} 
\end{center}

We wish to thank Jim Isenberg and Vince Moncrief for valuable discussions and 
for making available to us the manuscript of their forthcoming paper \cite{IM06}.  
We would also like to thank P. Chrusciel and I. Racz for useful discussions
and important suggestions.
We have benefited from workshops at Isaac Newton Institute (``Global Problems in 
Mathematical Relativity''), Oberwolfach (``Mathematical Aspects of General Relativity''), 
and at KITP UCSB (``Scanning New Horizons:GR Beyond 4 Dimensions''). 
This research was supported in part by NSF grant PHY 04-56619 
to the University of Chicago.

\bigskip 

\appendix 

\section{Ricci tensor in Gaussian null coordinates}\label{sect:A}

In this Appendix, we provide expressions for the Ricci tensor in a
Gaussian null coordinate system. As derived in section~\ref{sect:2}, 
in Gaussian null coordinates, the metric takes the form 
\ben
g_{ab} = 2 \left( 
                 \nabla_{(a} r - r \alpha \nabla_{(a}u - r \beta_{(a} 
           \right) \nabla_{b)}u 
       + \gamma_{ab} \,,  
\een
where the tensor fields $\beta_a$ and
$\gamma_{ab}$ are orthogonal to $n^a$ and $\ell^a$.
The horizon, $H$, corresponds to the surface $r=0$. We previously noted
that ${\gamma^a}_b$ is the orthogonal
projector onto the subspace of the tangent
space orthogonal to $n^a$ and $\ell^a$, and that when $r \beta_a \neq 0$,
it differs from the orthogonal 
projector, ${q^a}_b$, onto the surfaces $\Sigma(u,r)$. 
It is worth noting that in terms of the Gaussian null coordinate components 
of $\gamma_{ab}$, we have
$q^{ab} =
(\gamma^{-1})^{AB} (\partial /\partial x^A)^a (\partial/\partial x^B)^b$. 
It also is convenient
to introduce the non-orthogonal projector ${p^a}_b$, uniquely
defined by the conditions that ${p^a}_b n^b = {p^a}_b \ell^b = 0$ and that
${p^a}_b$ be the identity map on vectors that are tangent to $\Sigma(u,r)$. 
The relationship between ${p^a}_b$ and ${\gamma^a}_b$ is given by
\ben
{p^a}_b = -r \ell^a \beta_b + {\gamma^a}_b \,.  
\een
In terms of Gaussian null coordinates, we have
$p^a{}_b = (\partial/\partial x^A)^a (\dd x^A)_b$, from which it is easily 
seen that
$\pounds_n p^a{}_b = 0 = \pounds_\ell p^a{}_b$. It also is easily seen that
$q^{ac} \gamma_{cb} = p^a{}_b$ and that $p^a{}_b q^b{}_c = q^a{}_c$.

We define the derivative operator $D_c$ acting on a tensor 
field $T^{a_1 \dots a_r}{}_{b_1 \dots b_s}$ by the following prescription. 
First, we project the indices of the tensor field by $q^a{}_b$, then we 
apply the covariant derivative $\nabla_c$, and we then again project
all indices using $q^a{}_b$. For tensor fields intrinsic to $\Sigma$, this
corresponds to the derivative operator associated with the metric $q_{ab}$.
We denote the Riemann and Ricci tensors associated with $q_{ab}$ as
${\mathcal R}_{abc}{}^d$ and ${\mathcal R}_{ab}$. 

The Ricci tensor of $g_{ab}$ can then be written in the following form: 
\bena 
\label{nnR} 
 n^an^bR_{ab} &=&
 -\half q^{ab} \pounds_n\pounds_n \gamma_{ab} 
 +\quater q^{ca}q^{db}(\pounds_n \gamma_{ab})\pounds_n \gamma_{cd} 
 +\frac{1}{2} \alpha \: q^{ab}\pounds_n \gamma_{ab} 
\non \\
 &+&
 \frac{r}{2} \cdot 
 \Bigg[\;     
       4 \alpha \pounds_\ell \pounds_\ell \alpha 
       + 8 \alpha \pounds_\ell\alpha 
       + (\pounds_\ell\alpha) q^{ab}\pounds_n\gamma_{ab} 
\non \\ 
 && \qquad \,
      + q^{ab}\pounds_\ell\gamma_{ab} \cdot 
                \Big\{
                      -\pounds_n \alpha
                      -r q^{cd}\beta_c\pounds_n\beta_d 
\non \\
 && \qquad \qquad \qquad \qquad \,\,\: 
                      +(rq^{cd}\beta_c\beta_d+2\alpha)\pounds_\ell(r\alpha)
                      + rq^{cd}\beta_c D_d \alpha 
                \Big\}
\non \\ 
 && \qquad \, 
     + 2q^{ab}{D}_a  
          \left\{
             \beta_b\pounds_\ell(r\alpha) + {D}_b\alpha-\pounds_n\beta_b
          \right\} 
\non \\ 
 && \qquad \, 
     +q^{bc}\pounds_\ell(r\beta_c)\cdot
          \Big\{
                 (rq^{ef}\beta_e\beta_f+2\alpha) \pounds_\ell(r\beta_b) 
\non \\ 
 && \qquad \qquad \qquad \qquad \quad 
                - 4 D_b\alpha  
                + 2 \pounds_n\beta_b + 4rq^{ae}\beta_e {D}_{[a}\beta_{b]} 
          \Big\}
\non \\ && \qquad \,
     + 2(\pounds_\ell\alpha)\pounds_\ell(r^2q^{ab}\beta_a\beta_b) 
     + 4rq^{ab}\beta_a\beta_b \pounds_\ell\alpha 
     + 2 rq^{ab}\beta_a\beta_b \pounds_\ell \pounds_\ell \alpha 
\non \\ && \qquad \,  
     +2 q^{ab}\beta_a\pounds_\ell(r\beta_b)\cdot
      \left\{
             2 \pounds_\ell(r\alpha)
            - \half r q^{cd}\beta_c\pounds_\ell(r\beta_d)
      \right\} 
\non \\ 
 && \qquad \, 
     + 2r^{-1}\pounds_\ell \left\{ 
                                 r^2q^{ab}\beta_a(D_b\alpha-\pounds_n\beta_b)
                           \right\} 
     + 2r^{-1}\alpha \pounds_\ell(r^2 q^{ab}\beta_a\beta_b) 
 \Bigg] \,,
\eena

\bena 
\label{nlR} 
 n^a\ell^bR_{ab}
     &=& -2 \pounds_\ell\alpha 
         + \quater q^{ca}q^{db}(\pounds_n\gamma_{cd})\pounds_\ell\gamma_{ab} 
         -\half q^{ab} \pounds_\ell \pounds_n \gamma_{ab} 
         -\frac{1}{2} \alpha\: q^{ab} \pounds_\ell\gamma_{ab} 
         -\half q^{ab}\beta_a\beta_b  
\non \\
 &+& 
   \frac{r}{2}\cdot 
    \Bigg[  
          - 2\pounds_\ell \pounds_\ell\alpha 
          -\half q^{ab}\pounds_\ell\gamma_{ab} \cdot
           \left\{ 
                  2\pounds_\ell\alpha + q^{cd}\beta_c\pounds_\ell(r\beta_d) 
           \right\} 
\non \\
 && \qquad \, \, 
          - q^{ab}\beta_a\pounds_\ell\beta_b  
          - \pounds_\ell\{ q^{ab}\beta_a\pounds_\ell(r\beta_b)\} 
          - q^{ab}D_a(\pounds_\ell\beta_b) 
    \Bigg] \,, 
\eena 

\bena
\label{nAR} 
 n^bp^c{}_aR_{bc} &=& 
  - p^b{}_aD_b \alpha 
  + \half\pounds_n\beta_a  
  + \quater \beta_a q^{bc} \pounds_n \gamma_{bc} 
  - p^d{}_{[a}p^e{}_{b]}D_d(q^{bc}\pounds_n \gamma_{ce}) 
\non \\ 
&+& 
   \frac{r}{2}\cdot 
   \Bigg[\; 
         \half (q^{bc}\pounds_n\gamma_{bc})\pounds_\ell \beta_a 
         + \pounds_n\pounds_\ell\beta_a  
         + 2\alpha \pounds_\ell\beta_a  
\non \\ 
 && \qquad 
         + \pounds_\ell(r\beta_a) \cdot
                       \left\{ 
                              r^{-1}\pounds_\ell(r^2q^{bc}\beta_b\beta_c) 
                              + 2 \pounds_\ell\alpha  
                       \right\} 
\non \\
 && \qquad 
         - 2p^b{}_aD_b(\pounds_\ell\alpha) 
         + \pounds_\ell(q^{bc}\beta_b\pounds_n\gamma_{ca})  
         - 2r^{-1}\pounds_\ell \left( 
                                     r^2 q^{cd}\beta_cp^b{}_aD_{[b}\beta_{d]} 
                               \right) 
\non \\
 && \qquad 
         - \half q^{bc}\pounds_\ell\gamma_{bc}\cdot 
           \Big\{      
                  - (rq^{ef}\beta_e\beta_f+2\alpha)\pounds_\ell(r\beta_a) 
\non \\ 
&& \qquad \qquad \qquad \qquad \qquad    
                  + 2 p^d{}_aD_d\alpha 
                  - q^{bc}\beta_b\pounds_n \gamma_{ca} 
                  + 2r q^{ef}\beta_ep^d{}_aD_{[d}\beta_{f]} 
           \Big\}  
\non \\
 && \qquad 
         - 2 \pounds_\ell(\alpha \beta_a) 
         - 2 r(\pounds_\ell \alpha) \pounds_\ell \beta_a 
         + p^d{}_aD_b\left\{q^{bc}\beta_c\pounds_\ell(r\beta_d) \right\} 
\non \\ 
 && \qquad 
         - 2 p^b{}_a q^{cd}D_dD_{[b}\beta_{c]} 
         - q^{bc} (\pounds_\ell\beta_b) \pounds_n \gamma_{ca} 
\non \\
 && \qquad 
         - q^{bc}\pounds_\ell(r\beta_b)\cdot 
           \Big\{
                  (rq^{ef}\beta_e\beta_f+2\alpha) \pounds_\ell\gamma_{ca} 
                  + p^d{}_a D_c \beta_d 
\non \\ 
&& \qquad \qquad \qquad \qquad \qquad  
                  + \beta_c\pounds_\ell(r\beta_a) 
                  - rq^{ef}\beta_c\beta_f\pounds_\ell\gamma_{ea}  
           \Big\}     
\non \\
 && \qquad 
          + q^{bc}(\pounds_\ell\gamma_{ca})\cdot 
             \left\{  
                    2 \beta_b\pounds_\ell(r\alpha) 
                    + 2 D_b \alpha 
                    - \pounds_n \beta_b   
                    + 2r q^{de}\beta_eD_{[b}\beta_{d]}  
             \right\}  
   \Bigg]\,,  
\eena

\bena
\label{llR} 
 \ell^a \ell^b R_{ab} 
        &=& - \half q^{ab} \pounds_\ell \pounds_\ell \gamma_{ab} 
            + \quater q^{ca}q^{db}
              (\pounds_\ell\gamma_{ab})\pounds_\ell\gamma_{cd} \,,  
\eena

\bena 
\label{lAR} 
 \ell^bp^c{}_a R_{bc} 
      &=& 
         - \quater \beta_a q^{bc}\pounds_\ell \gamma_{bc} 
         - \pounds_\ell \beta_a 
         + \half q^{bc}\beta_c\pounds_\ell \gamma_{ab} 
         - p^d{}_{[a} p^e{}_{b]}D_d 
           \left(q^{bc}\pounds_\ell \gamma_{ce} \right) 
\non \\ 
 & + &  \frac{r}{2}\cdot 
        \Bigg[\;  
                - \pounds_\ell \pounds_\ell \beta_a 
               + \pounds_\ell 
                \left(q^{bc}\beta_c\pounds_\ell\gamma_{ab} \right) 
\non \\
&& \qquad \qquad 
                + \half (q^{cd}\pounds_\ell \gamma_{cd}) 
                    \left(
                          - \pounds_\ell\beta_a 
                          + q^{be}\beta_e\pounds_\ell \gamma_{ab}  
                    \right) 
        \Bigg] \,,  
\eena  

\bena 
\label{ABR}
p^c{}_ap^d{}_b R_{cd} 
        &=& 
         - \pounds_\ell\pounds_n \gamma_{ab} 
         - \alpha \pounds_\ell \gamma_{ab} 
         + p^c{}_ap^d{}_b {\cal R}_{cd} 
         - p^c{}_{(a} p^d{}_{b)} D_c\beta_d 
         - \half \beta_a\beta_b 
\non \\  
        &+& 
           q^{cd} \left(\pounds_\ell\gamma_{d(a}\right)\pounds_n \gamma_{b)c} 
         - \quater 
           \left\{ 
                   (q^{cd}\pounds_n \gamma_{cd})\pounds_\ell \gamma_{ab} 
                 + (q^{cd}\pounds_\ell \gamma_{cd})\pounds_n \gamma_{ab} 
           \right\} 
\non \\ 
 &+& \frac{r}{2} \cdot 
     \Bigg[ 
           - 2\alpha \pounds_\ell \pounds_\ell \gamma_{ab} 
           - p^e{}_ap^f{}_bD_c(q^{cd}\beta_d\pounds_\ell \gamma_{ef}) 
\non \\  
  && \qquad \, 
           - \half (q^{cd}\pounds_\ell\gamma_{cd}) 
             \left\{ 
                     (rq^{ef}\beta_e\beta_f+2\alpha)\pounds_\ell\gamma_{ab} 
                   + 2 p^e{}_{(a} p^f{}_{b)} D_e\beta_f 
             \right\} 
\non \\ 
  && \qquad \, 
           - 2(\pounds_\ell \alpha) \pounds_\ell \gamma_{ab} 
           - r^{-1}\{\pounds_\ell(r^2q^{ef}\beta_e\beta_f)\} 
                   \pounds_\ell \gamma_{ab} 
\non \\  
  && \qquad \, 
           - rq^{ef}\beta_e\beta_f \pounds_\ell \pounds_\ell \gamma_{ab} 
           - 2\pounds_\ell \{ p^c{}_{(a} p^d{}_{b)} D_c\beta_d \}  
\non \\ 
  && \qquad \, 
           - 2\beta_{(a}\pounds_\ell\beta_{b)}
           - r(\pounds_\ell\beta_a) \pounds_\ell\beta_b  
           - rq^{ce}q^{df}\beta_c\beta_d 
             (\pounds_\ell \gamma_{ae})\pounds_\ell\gamma_{bf} 
\non \\ 
  && \qquad \, 
           + 2q^{cd}\beta_d
             \left\{\pounds_\ell(r\beta_{(a}) \right\}\pounds_\ell\gamma_{b)c} 
           + 2p^e{}_{(a}p^f{}_{b)}q^{cd}\left(D_d \beta_e \right) 
             \pounds_\ell\gamma_{fc} 
\non \\ 
  && \qquad \, 
           + q^{cd} (r q^{ef}\beta_e\beta_f+2\alpha) 
                      (\pounds_\ell\gamma_{ca})\pounds_\ell\gamma_{db} \,  
     \Bigg] \,.        
\eena

\section{Maxwell equations in Gaussian null coordinates}\label{sect:B}
With the notation introduced in Appendix~\ref{sect:A} and 
the definitions (\ref{decomp}) and (\ref{VWU}), 
the Maxwell equations, $\nabla_a F^{ab}=0$, 
are equivalent to the following equations.
\bena 
0 &=& 
   \pounds_\ell S + \half S q^{ab} \pounds_\ell \gamma_{ab} 
   - q^{ab}\beta_aW_b - q^{ab}D_aW_b   
\non \\
 &-&  \frac{r}{2}\cdot
           \Big[\;   
                 2\pounds_\ell(q^{ab}\beta_aW_b)
                +q^{ab}q^{cd} \beta_aW_b \pounds_\ell \gamma_{cd} 
           \Big] \,,  
\label{maxwell:u}
 \\ 
0 &=& 
   \pounds_n S + \half S q^{ab} \pounds_n \gamma_{ab} + q^{ab}D_aV_b 
\non \\
  &-&  \frac{r}{2}\cdot
           \Big[\;   
                2 \pounds_n (q^{ab}\beta_aW_b) 
                + q^{ab} q^{cd}\beta_cW_d \pounds_n \gamma_{ab} 
\non \\
 && \qquad \, 
               - 2 q^{ab}D_a\left\{
                                 SV_b + 2\alpha W_b - q^{cd} \beta_dU_{bc} 
                            \right\} 
               - 4 rq^{ab}q^{cd}D_a(\beta_c \beta_{[d}W_{b]}) 
           \Big] \,, 
\label{maxwell:r}
 \\ 
0 &=& \half q^{ab}q^{cd} \left( 
                               W_b\pounds_n \gamma_{cd} 
                             + V_b \pounds_\ell \gamma_{cd} 
                         \right) 
      + \pounds_n (q^{ab}W_b) + \pounds_\ell (q^{ab}V_b)  
\non \\ 
 &+&   q^{ab}\left\{   
                   S\beta_b + 2\alpha W_b - q^{cd}\beta_d U_{bc} 
             \right\} 
\non \\ 
 &-& 
    \frac{r}{2} \cdot 
      \Big[
           -2 \pounds_\ell 
              \left\{
                 q^{ab}\left(
                             S\beta_b + 2\alpha W_b-q^{cd}\beta_dU_{bc} 
                       \right)   
              \right\}  
\non \\
  && \qquad 
            - 8 q^{ab}q^{cd}\beta_c\beta_{[d}W_{b]} 
            - 4 \pounds_\ell \left(
                                  q^{ab}q^{cd}\beta_c\beta_{[d}W_{b]}  
                             \right) 
\non \\
  && \qquad 
            - q^{cd}(\pounds_\ell \gamma_{cd})\cdot 
              q^{ab} \left\{
                            S\beta_b + q^{ef}\beta_e U_{fb}
                            + 2\alpha W_b - 2r q^{ef}\beta_e\beta_{[b}W_{f]} 
                     \right\}                   
      \Big] \,. 
\label{maxwell:A}
\eena 
The Bianchi identities, $\nabla_{[a} F_{bc]} = 0$, are given by 
\bena 
 \pounds_n W_a - \pounds_\ell V_a + p^c{}_aD_cS &=& 0 \,, 
\label{BI:urA}
\\
 \pounds_n U_{ab} - 2p^c{}_{[a} p^d{}_{b]} D_cV_d &=& 0 \,, 
\label{BI:uAB}
\\
 \pounds_\ell U_{ab} - 2p^c{}_{[a}p^d{}_{b]} D_cW_d &=& 0 \,, 
\label{BI:rAB}
\\ 
 p^d{}_{[a}p^e{}_bp^f{}_{c]}D_dU_{ef} &=& 0 \,.  
\label{BI:ABC}
\eena 
The stress tensor, $T_{ab} = F_{ac} F_b{}^c - (1/4) g_{ab} F^{cd} F_{cd}$, 
is given by 
\bena
n^an^bT_{ab} &=& q^{ab}V_aV_b 
              + r\cdot \Bigg[ \:     
                             2 q^{ab}\beta_aV_bS 
                            + (rq^{ab}\beta_a\beta_b+2\alpha)S^2 
                            + \half \alpha F^{cd}F_{cd} \:    
                       \Bigg] \,, 
\label{T:uu}
\\ 
n^a\ell^bT_{ab} 
 &=& -\half S^2 - \quater q^{ac}q^{bd}U_{ab}U_{cd} 
\non \\
 &+& r\cdot \Bigg[\:    
                    \alpha q^{cd}W_cW_d 
                  + q^{bc}\beta_c \left( 2 W_b S + q^{de}W_eU_{bd}\right) 
\non \\
 && \qquad \, 
            + \frac{r}{2}\left(q^{ab}q^{cd} - q^{ac}q^{bd}\right) 
                      \beta_a\beta_b W_cW_d 
         \: \Bigg] \,,  
\label{T:ur}
\\ 
n^bp^c{}_aT_{bc} &=& 
        - SV_a + q^{bc} U_{ab}V_c 
\non \\ 
     &+& r\cdot\Bigg[ \:  
                     q^{bc}\beta_c
                     \left(-W_aV_b + U_{ab}S \right) 
                     - \left(rq^{bc}\beta_b\beta_c + 2\alpha  \right)W_aS 
                     + \quater \beta_aF^{cd}F_{cd} 
            \: \Bigg] \,, 
\label{T:uA}
\\ 
  \ell^a\ell^bT_{ab} &=& q^{ab}W_aW_b \,, 
\label{T:rr}
\\
  \ell^bp^c{}_aT_{bc} &=& 
                        SW_a + q^{bc}U_{ab}W_c 
                         - rq^{bc}\beta_cW_aW_b \,, 
\label{T:rA}
\\
p^c{}_ap^d{}_bT_{cd} 
&=&   \half \gamma_{ab}S^2 + 2V_{(a}W_{b)} - \gamma_{ab}q^{cd}V_cW_d
      + q^{cd} \left\{ 
                      U_{ac}U_{bd} - \quater \gamma_{ab}q^{ef}U_{ce}U_{df} 
               \right\} 
\non \\ 
 &-& r\cdot \gamma_{ab}\cdot 
            \Bigg[ \:
                    \alpha q^{bc}W_bW_c                    
                  + q^{bc}\beta_c \left(W_bS+q^{de}W_eU_{bd} \right) 
 \non \\
  && \qquad \qquad   
                  + \frac{r}{2} \left( q^{cd}q^{ef}- q^{ce}q^{df}\right) 
                     \beta_c\beta_dW_eW_f
         \: \Bigg] \,, 
\label{T:AB}
\\
 T^c{}_c 
 &=& (n-4)\left\{\:
                 \half S^2 - q^{ab}V_aW_b - \quater q^{ab}q^{cd}U_{ac}U_{bd} 
        \:\right\}
\non \\
 &-& \frac{(n-4)}{2}\cdot r\cdot 
     \Bigg[ \: 
             2 \alpha q^{ab}W_aW_b 
           + 2 q^{ab}\beta_a \left( W_bS+ q^{cd}W_d U_{bc} \right) 
\non \\
 && \qquad \qquad \qquad          
          + r \left(q^{ab}q^{cd} - q^{ac}q^{bd}\right)\beta_a\beta_bW_cW_d     
   \: \Bigg] \,, 
\label{T:trace}
\eena 
where
\bena
 \quater F^{cd}F_{cd} 
  &=& q^{ab}V_aW_b -\half S^2 + \quater q^{ac}q^{bd}U_{ab}U_{cd} 
\non \\ 
  &+& r\cdot \Bigg[\:     
                     \alpha q^{cd}W_cW_d 
                   + q^{bc}\beta_c\left(
                                      W_bS+q^{de}W_eU_{bd} 
                                \right) 
\non \\ 
&& \qquad \, 
                   + \frac{r}{2}
                     \left(q^{ab}q^{cd} - q^{ac}q^{bd}\right)
                     \beta_a\beta_bW_cW_d 
          \: \Bigg]\,.       
\eena

\section{Analyticity of $f$ and $h$} \label{sect:C}

In this Appendix, we prove the following lemma, which establishes that 
if the spacetime $(M, g_{ab})$ and horizon $H$ are analytic, then 
the candidate Killing field $K^a$ also is analytic.

\paragraph{Lemma 3:}
If the spacetime as well as $\Sigma$ and $H$ are analytic, then the 
functions $f, h: \Sigma \to \mr$ given by eqs.~\eqref{fdef}
and~\eqref{psh} are real analytic. 

\medskip
\noindent
Proof: Consider first the function $p(x, \sigma)$ on $\Sigma \times \mr$
defined above in eq.~\eqref{pdef}. Let $x_0 \in \Sigma$ be fixed, 
and choose Riemannian normal coordinates $y^1, \dots, y^{n-2}$ around
$x_0$, so that the coordinate components $\gamma_{AB}(x_0)$ are given 
by $\delta_{AB}$. If $\alpha = (\alpha_1, \dots, \alpha_{n-2})
\in {\mathbb N}_0^{n-2}$ is a multi-index, we set 
$|\alpha| = \sum \alpha_i$, and $\alpha! = \prod \alpha_i!$, as well
as  
\ben
   \partial^\alpha = \frac{\partial^{|\alpha|}}{(\partial y^1)^{\alpha_1}
   \cdots (\partial y^{n-2})^{\alpha_{n-2}}} \, .
\een
We will show that, for $y$ in a sufficiently small ball around $x_0$, 
and for sufficiently large $\sigma$ we have the following estimate:
\ben\label{pap}
|\partial^\alpha p(y, \sigma)| \le \alpha! C R^{-|\alpha|} 
                                   \e^{-\sigma \kappa/2} \,, 
\een 
with $C$ and $R$ being some constants independent of $\alpha$. 
This implies that $f(y)$ has a convergent power series representation 
near $x_0$. Using Einstein's equation as in the proof of Lemma~1, we 
have 
\ben\label{pp}
\frac{\partial}{\partial y^A} p(y, \sigma) = [\beta_A(y) - (\hat
\phi_\sigma^* \beta)_A(y)] \, p(y, \sigma) \, . 
\een
Now we complexify $\Sigma$ and consider a complex multi-disk around 
$x_0$ of radius $R$. Then, using the multi-dimensional version of 
the Cauchy inequalities, we furthermore have the estimate 
\ben\label{paba}
|\partial^\alpha \beta_A(y)| \le \alpha! C R^{-|\alpha|} \,,  
\een
where $C$ is now taken as the supremum of $\beta_A$ in the complex multi-disk 
around $x_0$. 
Furthermore, because $\Sigma$ is compact, and because $\hat
\phi_\sigma$ is an isometry, we can find the same type of estimate 
in also for $\partial^\alpha (\hat \phi_\sigma^* \beta)_A(y)$
uniformly in $\sigma$ and $y$ in a ball around $x_0$. Finally,
from Lemma~1, we have the estimate
\ben
 |p(y,\sigma)| \le \e^{-\sigma (\kappa-\epsilon)} \,, 
\een
for arbitrary small $\epsilon > 0$ for any $y$ in a ball 
around $x_0$, and for sufficiently large $\sigma$. Applying now
further derivatives to eq.~\eqref{pp} and using the above 
estimates, we obtain the estimate~\eqref{pap}. 

In order to prove the analyticity of $h$, we must look at the 
explicit construction of that function given in the proof of Lemma~2. 
That construction shows that $h$ will be analytic, if we can show that 
the vector field $\beta^*(x)$ defined by the integral~\eqref{lim1} is 
analytic. This follows from the fact that the Taylor 
coefficient of the integrand $\partial^\alpha (\hat \phi_\tau^* \beta)_A(y)$
satisfies an estimate of the form~\eqref{paba}
uniformly in $\tau$ and $y$ in a ball around $x_0$. $\Box$ \hfill

\end{document}